\documentclass[a4paper,reqno]{amsart}
\usepackage{amssymb}
\usepackage{latexsym}
\usepackage{amsmath}
\usepackage{euscript}
\usepackage{graphics,xcolor}
\usepackage[all]{xy}
\usepackage[margin=3cm]{geometry}
\usepackage{MnSymbol} 

\newcommand{\be}{\begin{equation}}
\newcommand{\ee}{\end{equation}}
\newcommand{\ba}{\begin{eqnarray}}
\newcommand{\ea}{\end{eqnarray}}
\newcommand{\baa}{\begin{eqnarray*}}
\newcommand{\eaa}{\end{eqnarray*}}
\newcommand{\bb}{\begin{thebibliography}}
\newcommand{\eb}{\end{thebibliography}}

\newcounter{my}
\newcommand{\he}%
   {\stepcounter{equation}\setcounter{my}%
   {\value{equation}}\setcounter{equation}0%
   }%
\newcommand{\she}%
   {\setcounter{equation}{\value{my}}%
    }%

\newcommand{\clW}{W_{\!\mathrm{cl}}}

\theoremstyle{definition}

\numberwithin{equation}{section}

\begin{document}

\title{Rational Heun operators on $q$-linear grids}

\author{Satoshi Tsujimoto}
\author{Luc Vinet}
\author{Alexei Zhedanov}

\address{Graduate School of Informatics, Kyoto University,
Yoshida-Honmachi, Kyoto, Japan 606-8501}

\address{Centre de recherches \\ math\'ematiques,
Universit\'e de Montr\'eal, P.O. Box 6128, Centre-ville Station,
Montr\'eal (Qu\'ebec), H3C 3J7}

\address{Leonhard Euler International Mathematical Institute, Saint Petersburg, Russian Federation}



\begin{abstract}
Rational Heun operators on the $q-$linear grid are presented. They are second-order $q-$difference operators $W_q$ constructively defined from the requirement that they have a raising action on rational functions of type $[n/n]$, namely $W_q: [n/n] \rightarrow [n+1/n+1]$, with poles on  $q-$linear grids. It will be observed that these operators are related to one family of the Ruijsenaars-van Diejen-Takemura Hamiltonians. A distinguished subclass of $W_q$ called classical which shifts the pole structure while preserving the rational function type and a prescribed basis is also characterized. 
\end{abstract}

\maketitle

\section{Introduction}

Heun operators occupy a central position in the theory of exactly solvable models,
special functions, and integrable systems.
In their original differential form, Heun operators arise in the definition of the eponymous 
second-order Fuchsian differential equation with four regular singularities
\cite{ronveaux1995heun,kristensson2010second}.
They may be obtained by tridiagonalizing the hypergeometric operator, a construction
that explains their fundamental tridiagonal action on the Jacobi polynomials
\cite{erdelyi1944certain,grunbaum2017tridiagonalization}.

Over the years, Heun operators and equations have appeared in a wide variety of contexts.
Their polynomial solutions realize representation bases of $O(3)$
\cite{patera1973new}, they are related to Painlev\'e equations
\cite{slavyanov2015antiquantization}, they occur as Hamiltonians of Euler--Arnold
tops in external fields \cite{turbiner2016heun}, and they arise in the study of the
$BC_1$ Inozemtsev model and its degenerations
\cite{inozemtsev1989lax,oshima1995commuting,takemura2003heun}.
These connections have motivated the systematic search for discrete and
$q$-difference analogues of the Heun operator.

A striking observation is that the standard Heun operator  admits two equivalent
characterizations.
On the one hand, it is the most general second-order differential operators that
raises the degree of monomials by one.
On the other hand, it is the most general bilinear combination of the bispectral
operators associated with Jacobi polynomials
\cite{grunbaum2017tridiagonalization}.
This equivalence has been extended to a wide range of discrete and $q$-difference
settings, including Heun operators associated with Hahn, Racah, Bannai--Ito,
$q$-Jacobi, and Askey--Wilson polynomials
\cite{vinet2019heun,baseilhac2020q,bergeron2020heun,crampe2020heun,baseilhac2019heun}.
In several of these cases, the resulting $q$-Heun operators were identified with
one-particle Hamiltonians of the Ruijsenaars--van Diejen (RvD) hierarchy
\cite{ruijsenaars2004integrable,van1994integrability,takemura2017degenerations}.

A further step was taken in \cite{tsujimoto2023rational}, where the notion of Heun
operators was extended beyond the realm of orthogonal polynomials to that of
rational functions.
In that work, \textit{rational} Heun operators were defined as second-order difference
or $q$-difference operators acting as degree-raising maps on families of elementary
rational functions with prescribed poles.
With these poles taken on the Askey--Wilson grid, this approach led to so-called rational Heun operators 
that were seen to have a close link with degenerations of the Ruijsenaars--van Diejen Hamiltonians.
Furthermore, focusing on a subclass of these operators that preserves the rational function type, 
a generalized eigenvalue problem (GEVP) could be posited with the Wilson biorthogonal rational
functions of ${}_{10}\Phi_9$ type as solutions \cite{wilson1991orthogonal}, \cite{rahman1991biorthogonality}.

The present paper is a companion to \cite{tsujimoto2023rational} and
\cite{TVZ_tak}. In \cite{tsujimoto2023rational}, rational Heun operators
associated with the Askey--Wilson grid were introduced and shown to be related
to Takemura's degeneration $\mathcal A^{\langle1\rangle}$ of the
Ruijsenaars--van Diejen Hamiltonian. The purpose of the present work is to
investigate the corresponding theory on the $q$-linear grid.

Although the $q$-linear grid may be obtained as a degeneration of the
Askey--Wilson grid, the resulting operators possess a sufficiently distinctive
structure to warrant an independent treatment. We shall show that they admit a
natural characterization through a raising property on rational functions of
type $[n/n]$ and that they provide a realization of Takemura's degeneration
$\mathcal A^{\langle2\rangle}$.

Viewed together with the Heun operators associated with the Askey--Wilson,
big $q$-Jacobi and little $q$-Jacobi families, the results obtained here
complete the identification of the four one-particle Takemura degenerations
with Heun operators and rational Heun operators. This correspondence provides
further evidence of the deep connection between Heun operators,
bispectrality and the Ruijsenaars--van Diejen hierarchy.

The paper is organized as follows. In Section~2 we introduce the
$q$-linear pole configuration and formulate the raising prescription that
defines the rational Heun operators considered here. In Section~3 we derive
explicitly the corresponding second-order $q$-difference operator and show
that, after a suitable parametrization and gauge transformation, it gives a
Heun-theoretic realization of Takemura's degeneration
$\mathcal A^{\langle2\rangle}$ of the one-particle Ruijsenaars--van Diejen
Hamiltonian. Section~4 is devoted to the classical subclass, characterized by
the preservation of the rational type under a shift of the pole sequence.
Concluding remarks, including the relation with the Askey--Wilson case and
perspectives on Gupta--Masson rational functions, are given in Section~5.

.

\section{Rational Heun operators on the $q$-linear grid}

\subsection{Setup and construction principle}

Throughout this paper we work with second-order $q$-difference operators acting on functions of a complex variable $z$,
\begin{equation}
\label{eq:genericW}
W = A_1(z)\,\mathcal T_q + A_2(z)\,\mathcal T_{q^{-1}} + A_0(z)\,\mathcal I,
\end{equation}
where $\mathcal T_q f(z)=f(qz)$, $\mathcal T_{q^{-1}} f(z)=f(z/q)$, and $\mathcal I$ denotes the identity operator.
The coefficient functions $A_1(z)$, $A_2(z)$ and $A_0(z)$ are assumed to be rational functions of $z$.

We are interested in the action of such operators on families of elementary rational functions with prescribed poles lying on a $q$-linear grid.
In the present setting, this grid is taken to be
\begin{equation}
\label{eq:qlineargrid}
x_n = \alpha q^n , 
\qquad n\in\mathbb Z_{\ge 0},
\end{equation}
where $\alpha$ 
is a nonzero complex parameter.
We take $W$ to act on rational functions defined as linear combinations of the  elementary functions of the form 
\begin{equation}
\label{eq:elementaryRF}
\frac{z-x_0}{z-x_n}.
\end{equation}
where the parameters $x_n$ are prescribed. 

As in the polynomial and Askey--Wilson settings \cite{vinet2019heun,baseilhac2019heun,tsujimoto2023rational},
the Heun operator of interest will be defined through a raising property. 
This prescription will be stated explicitly in Subsection~2.2 for the situation of interest here.

The construction principle to be used can be summarized as follows.
We fix a pole configuration, a raising operation, and further assume that $W$ is of the general second-order form \eqref{eq:genericW}.
Requiring that $W$ acts in the prescribed way on the first few elementary rational functions—typically those with poles at $x_1,x_2$ in addition to the constant $1$ - yields a finite system of algebraic constraints on the coefficient functions $A_1(z)$, $A_2(z)$ and $A_0(z)$.
Solving these constraints determines $W$ up to a finite number of free parameters.
Once an operator $W$ has been constructed in this way, we verify by direct computation that its action on a general rational function with poles on the prescribed $q$-linear grid indeed produces a rational function of the required type.
In the case considered here, this verification confirms that the finite set of constraints imposed at the level of elementary rational functions is sufficient. In practical terms, the construction proceeds in three steps.
First, one imposes the raising prescription on the constant function and on the first elementary rational functions of the pole sequence. This yields a finite system of algebraic constraints on the coefficients of the generic second-order $q$-difference operator \eqref{eq:genericW}. Second, these constraints are solved explicitly, leading to a finite-parameter family of operators. Finally, one verifies that the resulting operator acts correctly on an arbitrary element of the rational basis, thereby establishing the sufficiency of the constraints. This procedure may be viewed as the rational analogue of the characterization of ordinary Heun operators through their degree-raising property.

\subsection{Raising prescriptions on the $q$-linear grid}

We now specify the pole-raising prescription that characterizes the rational $q$-Heun operators considered in this paper by describing how the second-order $q$-difference operator acts on families of elementary rational functions with poles lying on a $q$-linear grid.

\medskip
Let $\{x_n\}_{n\ge0}$ be a $q$-linear sequence as in \eqref{eq:qlineargrid}, and assume that the points $x_n$ are distinct for $n =0,1,2,\dots$.
We consider rational functions that admit a partial fraction decomposition of the form
\begin{equation}
\label{eq:Rnm}
R_{n}(z)
= a_{n,0} + 
\sum_{k=1}^{n} a_{n,k}\frac{z-x_0}{z-x_k}
\;\in\;
\mathrm{Span}\!\left\{1, 
\frac{z-x_0}{z-x_1},\ldots,\frac{z-x_0}{z-x_{n}}
\right\},
\end{equation}
where the coefficients $a_{n,0}$ and $a_{n,k}$ are arbitrary complex numbers.
Such a rational function will be said to be of type $[n/n]$.
We consider the raising condition 
\begin{equation}
\label{eq:raising2}
W R_{n}(z)
= \tilde a_{n,0} + 
\sum_{k=1}^{n+1} \tilde a_{n,k}\frac{z-x_0}{z-x_k}
\;\in\;
\mathrm{Span}\!\left\{1, 
\frac{z-x_0}{z-x_1},\ldots,\frac{z-x_0}{z-x_{n+1}}
\right\}.
\end{equation}
This prescription defines a rational Heun operator that maps functions of type $[n/n]$ to functions of type $[n+1/n+1]$.

\section{Explicit construction on the $q$-linear grid}
\label{sec:explicit}

Implementing the construction principle described in Section~2, one imposes the raising condition \eqref{eq:raising2} on the constant function and on the first two elementary rational functions of the pole sequence. Solving the resulting algebraic constraints leads to the family of rational $q$-Heun operators described below.

\subsection{Explicit expression}

The resulting rational $q$-Heun operator takes the form
\begin{equation}
\label{eq:Wq1_def}
W_{\!q} = A_{q,1}(z)\,\mathcal T_q
           + A_{q,2}(z)\,\mathcal T_{q^{-1}}
           + A_{q,0}(z)\,\mathcal I,
\end{equation}
with coefficient functions
\begin{align}
\label{eq:Aq1_single}
A_{q,1}(z) &= \frac{q\,(z-\alpha)\,Q_3(z)}{z^2},\\[2mm]
\label{eq:Aq2_single}
A_{q,2}(z) &= \frac{Q_5(z)}{z^2\,(z-\alpha q)},\\[2mm]
\label{eq:Aq0_single}
A_{q,0}(z) &= c_0 + c_1 z + c_2 z^2 + \frac{c_3}{z} + \frac{c_4}{z^2}.
\end{align}
Here $Q_3$ and $Q_5$ are polynomials of degrees $3$ and $5$,
\begin{equation}
\label{eq:Q3Q5_single}
Q_3(z)=\sum_{j=0}^3 \rho_{1,j} z^j,\qquad
Q_5(z)=\sum_{j=0}^5 \rho_{2,j} z^j,
\end{equation}
with the constraints
\begin{equation}
\label{eq:rho_constraints_single}
\rho_{2,5}=\rho_{1,3},\qquad
\rho_{2,0}=\alpha^2 q^3\,\rho_{1,0}.
\end{equation}
The coefficients in $A_{q,0}$ are given by 
\begin{align}
\label{eq:c_single}
c_1&=-\bigl(q\rho_{1,2}+\rho_{2,4}\bigr),\nonumber\\
c_2&=-(1+q)\rho_{1,3},\nonumber\\
c_3&=\alpha q\,\rho_{1,1}+(\alpha q)^{-1}\rho_{2,1},\nonumber\\
c_4&=\alpha\,q(1+q)\rho_{1,0}.
\end{align}

\subsection{Sufficiency condition}
The operator $W_{q}$ was obtained by solving the necessary conditions imposed by the prescribed actions on the constant function and the two elementary rational functions
\begin{equation}
    1, \quad\frac{z-x_0}{z-x_1},\quad\frac{z-x_0}{z-x_2}
\end{equation}
with the poles on $q$-linear grids as per \eqref{eq:qlineargrid}.
These conditions uniquely determine the functions $A_{q,1}(z)$, $A_{q,2}(z)$ and $A_{q,0}(z)$ entering in the defining form \eqref{eq:Wq1_def} of $W_{q}$. Their sufficiency follows from the fact that the actions on the generic elementary rational functions of the pole sequence can be explicitly computed and read: 
\begin{align}
    W_q\left(\frac{z-x_0}{z-x_n}\right)
    =\tilde a_{n,0} 
    +
    \tilde a_{n,n-1} \frac{z-x_0}{z-x_{n-1}}
    +
    \tilde a_{n,n} \frac{z-x_0}{z-x_n}
    +
    \tilde a_{n,n+1} \frac{z-x_0}{z- x_{n+1}},
\end{align}
where
\begin{align}
& \tilde{a}_{n,0} = \frac{Q_5(\alpha)}{\alpha^3 (1-q^{n+1})},\\
& \tilde{a}_{n,n-1} =  \frac{(q^n-1) Q_3(\alpha q^{n-1})}{\alpha q^{2n-2}}, \\
& \tilde{a}_{n,n} = c_0 + \frac{\rho_{2,1}}{\alpha^2q^{n+1}}+\frac{(1+q)\rho_{1,0}}{\alpha q^{2n-1}}+\frac{\rho_{1,1}}{q^{n-1}}  - \alpha q^{n}(q \rho_{1,2} + \alpha q^n (1+q)\rho_{1,3} + \rho_{2,4}),\\
& \tilde{a}_{n,n+1} = \frac{Q_5(\alpha q^{n+1})}{\alpha^3 q^{2n+2} (q^{n+1}-1)} .
\end{align}
This establishes that the constraints obtained from the first elementary rational functions are sufficient and hence that the operator $W_q$ satisfies the raising property \eqref{eq:raising2}.

\subsection{Symmetric $\varepsilon$-parametrization}
\label{subsubsec:epsparam_single}

It is convenient to reparametrize the coefficients of $W_q$ in terms of
elementary symmetric functions of auxiliary parameters
$\varepsilon_1,\ldots,\varepsilon_8$.
For variables $u_1,\ldots,u_N$, let $\sigma_k(u_1,\ldots,u_N)$ denote the
$k$-th elementary symmetric polynomial, namely
\[
\sigma_k(u_1,\ldots,u_N)
=
\sum_{1\le i_1<\cdots<i_k\le N}
u_{i_1}\cdots u_{i_k}.
\]
We introduce the $\varepsilon$-variables by
\begin{align}
\rho_{1,0}&=-q^{-\frac32} \rho_{1,3}\,\sigma_3(\varepsilon_1,\varepsilon_2,\varepsilon_3),\\
\rho_{1,1}&={\hspace{6pt}}q^{-1}\,\rho_{1,3}\,\sigma_2(\varepsilon_1,\varepsilon_2,\varepsilon_3),\\
\rho_{1,2}&=-q^{-\frac12}\rho_{1,3}\,\sigma_1(\varepsilon_1,\varepsilon_2,\varepsilon_3),\\
\rho_{2,0}&=-q^{\frac52}\rho_{1,3}\, \sigma_5(\varepsilon_4,\varepsilon_5,\varepsilon_6,\varepsilon_7,\varepsilon_8),\\
\rho_{2,1}&={\hspace{6pt}} q^2{\hspace{2pt}}\rho_{1,3}\,\sigma_4(\varepsilon_4,\varepsilon_5,\varepsilon_6,\varepsilon_7,\varepsilon_8),\\
\rho_{2,2}&=-q^{\frac32}\rho_{1,3}\, \sigma_3(\varepsilon_4,\varepsilon_5,\varepsilon_6,\varepsilon_7,\varepsilon_8),\\
\rho_{2,3}&={\hspace{6pt}} q{\hspace{6pt}}\rho_{1,3}\,\sigma_2(\varepsilon_4,\varepsilon_5,\varepsilon_6,\varepsilon_7,\varepsilon_8),\\
\rho_{2,4}&=-q^{\frac12} \, \rho_{1,3}\,\sigma_1(\varepsilon_4,\varepsilon_5,\varepsilon_6,\varepsilon_7,\varepsilon_8),
\end{align}
together with the constraint
\begin{equation}
\alpha^2
=
q\,\varepsilon_1^{-1}\varepsilon_2^{-1}\varepsilon_3^{-1}
\,\varepsilon_4\varepsilon_5\varepsilon_6\varepsilon_7\varepsilon_8
\end{equation}
from $\rho_{2,0}=\alpha^2 q^3\,\rho_{1,0}$.

If we take 
\begin{equation}
\alpha
= q^{\frac12}\,(\varepsilon_1 \varepsilon_2 \varepsilon_3)^{-\frac12}
(\varepsilon_4\varepsilon_5\varepsilon_6\varepsilon_7\varepsilon_8)^{\frac12}
\end{equation}
and the parametrization given above, the coefficient functions of $W_q$ can be written
in a factorized form:
\begin{align}
A_{q,1}(z)
&=\rho_{1,3}
 \frac{ (z\, (\varepsilon_1\varepsilon_2\varepsilon_3)^{\frac12}- q^{\frac12}\,(\varepsilon_4\varepsilon_5\varepsilon_6\varepsilon_7\varepsilon_8)^{\frac12})
\prod_{j=1}^3(q^{\frac12}z-\varepsilon_j)}{q^{\frac12}\, (\varepsilon_1\varepsilon_2\varepsilon_3)^{\frac12}\, z^2},\\
A_{q,2}(z)
&=\rho_{1,3}
\frac{ (\varepsilon_1\varepsilon_2\varepsilon_3)^{\frac12} \prod_{j=4}^8(z-q^{\frac12} \varepsilon_j)}
{z^2 (z\,(\varepsilon_1\varepsilon_2\varepsilon_3)^{\frac12} - q^{\frac32}(\varepsilon_4\varepsilon_5\varepsilon_6\varepsilon_7\varepsilon_8)^{\frac12})},\\
A_{q,0}(z)
&=
c_0
+ \rho_{1,3}\left( q^{\frac12} z\sum_{k=1}^8\varepsilon_k
-(1+q) z^2
+ q^{\frac12} z^{-1}\Bigl(\prod_{\ell=1}^{8}\varepsilon_{\ell}^{\frac12}\Bigr)\sum_{k=1}^8\varepsilon_k^{-1}
-(1+q)z^{-2}\prod_{\ell=1}^8\varepsilon_\ell^{\frac12} \right).
\end{align}
The corresponding pole sequence is furthermore
\begin{equation}
x_n = q^{n+\frac12}(\varepsilon_1 \varepsilon_2 \varepsilon_3)^{-\frac12}
(\varepsilon_4\varepsilon_5\varepsilon_6\varepsilon_7\varepsilon_8)^{\frac12} .
\end{equation}

\subsection{Connection to the Takemura Hamiltonian $\mathcal A^{\langle2\rangle}$}

The operator $W_q$ can be mapped to Takemura’s degeneration $\mathcal A^{\langle2\rangle}$ \cite{takemura2017degenerations},
\begin{align}
    \mathcal A^{\langle2\rangle} = V_1(z) {\mathcal T}_q 
    +
  V_2(z) {\mathcal T}_{q^{-1}}
  + V_0(z) {\mathcal I},
\end{align} 
where
\begin{align}
& V_1(z) = z^{-2} \prod_{k=1}^{4}(1-z \epsilon_k q^{\frac12})\cdot \prod_{k=5}^8 \epsilon_k,\\
& V_2(z) = q^{-1} z^2 \prod_{k=5}^{8}(1-z^{-1} \epsilon_k q^{\frac12}),\\
& V_0(z) =
\prod_{k=5}^{8} \epsilon_k
\left(\left(\sum_{\ell =1}^4 \epsilon_{\ell}+ \sum_{\ell =5}^8 \epsilon_{\ell}^{-1} \right) q^{\frac12} z^{-1} - (1+q) z^{-2}\right)
\nonumber \\
&\qquad \qquad +
\prod_{k=1}^{8} \epsilon_k^{\frac12}
\left(\left(\sum_{\ell =1}^4 \epsilon_{\ell}^{-1}+ \sum_{\ell =5}^8 \epsilon_{\ell} \right) q^{\frac12} z - (1+q) z^2\right)
\end{align}
by suitable gauge transformations 
\begin{align}
     \hat W_q (\varepsilon_1,\varepsilon_2,\varepsilon_3,\varepsilon_4,\varepsilon_5,\varepsilon_6,\varepsilon_7,\varepsilon_8,\rho_{1,3}) = \xi(z) W_q \,\xi(z)^{-1},
\end{align}
where
\begin{align}
    \xi(z q) = \frac{1-q^{-\frac12} z\, (\varepsilon_{1} \varepsilon_{2} \varepsilon_{3})^{\frac12}(\varepsilon_{4} \varepsilon_{5} \varepsilon_{6} \varepsilon_{7} \varepsilon_{8})^{-\frac12}}{1-q^{\frac12}z\, \varepsilon_{4}^{-1}} \,\xi(z) 
\end{align}
and
specializations of parameters as
\begin{align}
\mathcal A^{\langle2\rangle} = \hat W_q (\epsilon_{1}^{-1},\epsilon_{2}^{-1},\epsilon_{3}^{-1},\epsilon_{4}^{-1},\epsilon_5,\epsilon_6,\epsilon_7,\epsilon_8, (\epsilon_{1}\epsilon_{2}\epsilon_{3}\epsilon_{4}\epsilon_{5}\epsilon_{6}\epsilon_{7}\epsilon_{8})^{\frac12}).
\end{align}

The Heun operators associated with the big and little $q$-Jacobi polynomials on the $q$-linear grid were obtained in \cite{baseilhac2020q} by both the raising and tridiagonalization approaches. They are identified with Takemura’s degenerations $\mathcal A^{\langle 3 \rangle}$ and $\mathcal A^{\langle 4 \rangle}$ of the one-particle Ruijsenaars–van Diejen Hamiltonian. In light of the connection of the rational Heun operators associated to the Askey grid with the Takemura Hamiltonian $\mathcal A^{\langle 1 \rangle}$ initiated in \cite{tsujimoto2023rational} and completed in \cite{TVZ_tak}, the results recorded here thus complete the identification of each of these four one-particle degenerations of the Ruijsenaars–van Diejen Hamiltonian with Heun operators. This observation places the present construction within a broader
Heun--Ruijsenaars--van Diejen correspondence that now encompasses all four one-particle Takemura degenerations.

\section{Classical rational Heun operators on the $q$-linear grid}
\label{sec:classical}

In \cite{tsujimoto2023rational}, a distinguished subclass of rational Heun operators
was identified by imposing a classical condition.
This condition requires that the operator preserves the pole type, namely that
it sends rational functions of type $[n/n]$ with poles at $x_1,\ldots,x_n$
to rational functions of the same type with poles at $x_2,\ldots,x_{n+1}$.

More precisely, let $R_n(z)$ be a rational function of type $[n/n]$, given by \eqref{eq:Rnm},
with ordered poles at $x_1,\ldots,x_n$.
For a general rational Heun operator, its action on $R_n(z)$ is described by \eqref{eq:raising2}.
By contrast, for a classical Heun operator $\clW$, one imposes the condition that
the first pole $x_1$ does not appear in the resulting expression:
\begin{align}
\clW R_n(z) 
\;\in\;
\mathrm{Span}\!\left\{1, 
\frac{z-x_0}{z-x_2},\ldots,\frac{z-x_0}{z-x_{n+1}}
\right\}.
\end{align}
A rational Heun operator satisfying this condition is called a classical Heun operator.
As a consequence, $\clW R_n(z)$ is again a rational function of type $[n/n]$ with shifted poles at $x_2,\ldots,x_{n+1}$.

For the rational Heun operator $W_{\!q}$ in \eqref{eq:Wq1_def} to be a classical Heun operator, it is necessary that 
\begin{align}
\tilde a_{0,1}=0,\qquad \tilde a_{1,1}=0,\qquad \tilde a_{2,1}=0   
\end{align}
for $n=0,1,2$.
These conditions lead to
\begin{align}
\label{classical_rho:cond1}
  \rho_{1,3} &=-\frac{\rho_{1,2}}{\alpha  q}-\frac{\rho_{1,1}}{\alpha^{2} q^{2}}-\frac{\rho_{1,0}}{\alpha^{3} q^{3}},\\
\label{classical_rho:cond2}
   \rho_{2,4}+\alpha q \rho_{1,3} &=-\frac{\rho_{2,3}}{\alpha  q}-\frac{\rho_{2,2}+q \rho_{1,0}}{\alpha^{2} q^{2}}
   -\frac{\rho_{2,1}}{\alpha^{3} q^{3}},\\
\label{classical_rho:cond3}
   c_0 - \alpha  q \rho_{2,4}-\alpha^{2} q^{2} \rho_{1,3}&=-\rho_{1,1} \left(q +1\right)-\frac{\rho_{1,0} \left(2 q +1\right)}{q \alpha}-\frac{\rho_{2,1}}{\alpha^{2} q^{2}}.
\end{align}
The classical conditions \eqref{classical_rho:cond1}-\eqref{classical_rho:cond3} can be expressed in terms of the $\varepsilon$-variables introduced in Section \ref{subsubsec:epsparam_single}. Solving these equations for $\varepsilon_7$, $\varepsilon_8$, and  $c_0$
yields several solutions related by  symmetries and sign choices. For definiteness, we choose the following representative:
\begin{align}
    & \varepsilon_7 = q^{-\frac12}\varepsilon_1^{\frac12},\\
    & \varepsilon_8 = q^{-\frac32} \varepsilon_1(\varepsilon_2\varepsilon_3)^{\frac12} (\varepsilon_4\varepsilon_5\varepsilon_6)^{-\frac12},\\
    &c_0 = \rho_{1,3}\left(-\frac{q^{2} \varepsilon_{4} \varepsilon_{5} \varepsilon_{6}}{\varepsilon_{1}}-\frac{\varepsilon_{1}^{3} \varepsilon_{2} \varepsilon_{3}}{q^{3} \varepsilon_{4} \varepsilon_{5} \varepsilon_{6}} - \varepsilon_1 \sum_{j=2}^6 \left(\varepsilon_{j}+\frac{\varepsilon_{2} \varepsilon_{3}}{q \varepsilon_{j}}\right)\right).
\end{align}
Although this system admits several solutions for $c_0$,  $\varepsilon_7$, and $\varepsilon_8$, the apparent multiplicity is due only to the symmetry among $\varepsilon_1,\varepsilon_2,\varepsilon_3$. Hence they are equivalent up to permutations of $\varepsilon_1,\varepsilon_2,\varepsilon_3$, and we record only one representative solution.

Thus we obtain the classical Heun operator on the $q$-linear grid in the form 
\begin{equation}
\label{eq:CHq}
 \clW
 =
 B_1(z)\,(\mathcal T_q-\mathcal I)
 +
 B_2(z)\,(\mathcal T_{q^{-1}}-\mathcal I)
 +
 B_0\,\mathcal I,
\end{equation}
where 
\begin{align}
\label{def:classicla_B1}
B_1(z)
&=\rho_{1,3}
\dfrac{
(q^{\frac32} z -\varepsilon_1)
(q^{\frac12} z - \varepsilon_1)
(q^{\frac12} z - \varepsilon_2)
(q^{\frac12} z - \varepsilon_3)
}{q^2 z^2},\\
\label{def:classicla_B2}
B_2(z)
&=\rho_{1,3}
\dfrac{
(q^{\frac52} z\,\varepsilon_4\varepsilon_5\varepsilon_6-\varepsilon_1^2\varepsilon_2\varepsilon_3)
(z-q^{\frac12}\varepsilon_4)
(z-q^{\frac12}\varepsilon_5)
(z-q^{\frac12}\varepsilon_6)
}{q^{\frac52}\, z^2\, \varepsilon_4\varepsilon_5\varepsilon_6},\\
\label{def:classicla_B0}
B_0
&=
\rho_{1,3} \dfrac{
(q^2\varepsilon_4\varepsilon_5\varepsilon_6-\varepsilon_1\varepsilon_2\varepsilon_3)
(\varepsilon_1-q\varepsilon_4)
(\varepsilon_1-q\varepsilon_5)
(\varepsilon_1-q\varepsilon_6)
}{q^3\varepsilon_1 \varepsilon_4\varepsilon_5\varepsilon_6 }.
\end{align}
The corresponding pole sequence is $x_n = q^{n-\frac32}\varepsilon_1$ for $k=0,1,2,\dots$. 

The operator \eqref{eq:CHq} provides the natural rational analogue on the
$q$-linear grid of the classical Heun operators encountered in the polynomial
setting. Its defining property is the preservation of the rational function
type together with a shift of the pole configuration along the grid. As in the
Askey--Wilson case studied in \cite{tsujimoto2023rational}, this feature makes
it possible to formulate generalized eigenvalue problems whose solutions are
expected to be expressed in terms of distinguished families of biorthogonal
rational functions. We return to this point in the concluding remarks.

\section{Concluding remarks}
\label{sec:conclusion}

In this paper we have constructed rational Heun operators on the $q$-linear grid from a pole-raising prescription acting on rational functions of type $[n/n]$. Explicit expressions for these operators were obtained and shown to admit a factorized parametrization in terms of elementary symmetric functions.

A principal result is the identification of these rational Heun operators with Takemura's degeneration $\mathcal A^{\langle2\rangle}$ of the one-particle Ruijsenaars--van Diejen Hamiltonian. Together with the previously established correspondences involving the Askey--Wilson, big $q$-Jacobi and little $q$-Jacobi Heun operators, this completes the Heun-theoretic interpretation of the four one-particle Takemura degenerations.

We have furthermore characterized the classical rational Heun operators associated with the $q$-linear grid. These operators preserve the rational function type while shifting the pole configuration and provide the natural framework for the formulation of generalized eigenvalue problems. This characterization is obtained from the raising-operator perspective.
One of the long-term goals of the theory is to relate this constructive
approach to the algebraic and bispectral descriptions of rational special
functions, in the same way that ordinary Heun operators can be characterized
both through raising properties and through bilinear combinations of
bispectral operators.

The $q$-linear grid considered in this paper may be obtained as a degeneration
of the Askey--Wilson grid
\[
x(s)=a q^s+a^{-1}q^{-s}.
\]
Indeed, if one lets
\[
a\rightarrow\infty
\]
while keeping
\[
aq^s=\alpha q^n
\]
finite, the contribution from the second branch of the lattice disappears,
\[
a^{-1}q^{-s}\rightarrow 0,
\]
and the Askey--Wilson grid collapses to the $q$-linear sequence
\[
x_n=\alpha q^n.
\]
Under the same scaling, the elementary rational functions with poles on the
Askey--Wilson grid degenerate to the rational functions considered here.
Consequently, the rational Heun operators introduced in
\cite{tsujimoto2023rational} admit corresponding $q$-linear limits.
The results obtained in the present work can therefore be recovered from the
Askey--Wilson framework by degeneration. At the operator level, the rational Heun operators associated with the
Askey--Wilson grid degenerate to the rational Heun operators constructed
here, while the correspondence with the Ruijsenaars--van
Diejen--Takemura Hamiltonians is preserved. Nevertheless, the $q$-linear case
possesses its own distinctive features and admits a considerably simpler
description. In particular, the pole structure, the explicit parametrization
of the rational Heun operators and the connection with Takemura's
degeneration $\mathcal A^{\langle2\rangle}$ become especially transparent in
this setting.

A natural continuation of this work is the analysis of the generalized eigenvalue problems associated with the classical rational Heun operators introduced here. Their solutions are expected to be expressible in terms of Gupta--Masson biorthogonal rational functions. Another important direction concerns the algebraic interpretation of rational Heun operators. The present work was developed from the raising-operator perspective, which
provides a direct and constructive characterization of rational Heun
operators. An important challenge is now to connect this approach with the
emerging algebraic description of biorthogonal rational functions.

For ordinary Heun operators, the raising-operator approach and the
algebraic description based on bispectral operators ultimately coincide. Recent developments involving Leonard trios, meta algebras and rational
bispectrality strongly suggest that the ingredients of such a picture are now
available in the rational setting. The results obtained here provide additional evidence in this
direction. It is natural to expect that the rational Heun operators associated
with the various grids can eventually be characterized both by raising
properties and by algebraic relations, thereby leading to a unified framework
for rational Heun operators, rational bispectrality and the associated
families of biorthogonal rational functions.

\section*{Acknowledgments}
This work has been sponsored by a Québec-Kyoto cooperation grant from the Ministère des Relations Internationales et de la Francophonie of the Quebec Government.
The research of ST is supported by JSPS KAKENHI (Grant Number 24K00528). LV is funded in part through a discovery grant of the Natural Sciences and Engineering Research Council (NSERC) of Canada. 
 AZ is supported by the Ministry of Science and Higher Education of the Russian Federation (agreement no. 075–15–2025–343).

\section*{Conflict of interest}
On behalf of all authors, the corresponding author states that there is no conflict of interest.

\section*{Data availability}
This manuscript has no associated data.

\bibliographystyle{unsrt} 
\bibliography{ref_qlin.bib} 

\end{document}